\documentclass{cflowproc}

\usepackage{natbib}

\begin{document}


\shortauthors{KEMPNER ET AL.}     
\shorttitle{CLASSIFICATION OF RADIO SOURCES IN CLUSTERS} 

\title{Conference Note: A Taxonomy of Extended Radio Sources in Clusters of Galaxies}

\author{Joshua C. Kempner,\affilmark{1} 
Elizabeth L. Blanton,\affilmark{2}         
Tracy E. Clarke,\affilmark{2}
Torsten A. En{\ss}lin,\affilmark{3}
Melanie Johnston-Hollitt,\affilmark{4}
and Lawrence Rudnick\affilmark{5}}

\affil{1}{Harvard-Smithsonian Center for Astrophysics, 60 Garden St., MS-67, Cambridge, MA 02138}
\and
\affil{2}{University of Virginia, P.O. Box 3818, Charlottesville, VA 22903}
\and
\affil{3}{Max-Planck-Institut fuer Astrophysik, Karl-Schwarzschild-Str.\ 1, Postfach 1317, Garching D-85741, Germany}
\and
\affil{4}{Leiden Observatory, P.O. Box 9513, Leiden NL-2300 RA, The Netherlands}
\and
\affil{5}{University of Minnesota, 116 Church St.\ SE, Minneapolis, MN 55455}


\begin{abstract}
At the request of the conference attendees, we have compiled a
classification of extended radio sources in clusters.  These range from
scales of tens of parsecs to over a megaparsec in scale, and include both
sources associated with AGN and sources thought to derive from the electron
population in the ionized ICM.  We pay special attention to distinguishing
between the types of AGN in the cores of cooling flow clusters and between
the multiple classes of objects referred to over the years as ``radio
relics.''  We suggest new names based on physical arguments for some of
these classes of objects where their commonly used names are inappropriate
or confusing.
\end{abstract}


\section{Introduction}
\label{Confnote:intro}

This conference note was inspired by the frequent lamentation during
discussions of radio halos and relics that the nomenclature for these
sources is confusing.  ``We need a new name for relics'' has become a
common refrain, since three physically distinct sources are all referred to
in the literature as ``radio relics,'' and at least one of them is not a
relic of anything.  The phenomenological approach used to classify these
diffuse sources has produced this confusion, whereas a classification
scheme based on physical properties of the sources would suffer no such
drawback.  To make matters worse, the radio galaxies at the centers of
clusters are often referred to by a Fanaroff-Reilly type when in fact
neither an FR~I nor FR~II classification is appropriate.  This short paper,
then, is intended to clear up this confusion to the extent possible.  The
confusion will only go away, however, if the classifications described
herein are adopted by the community at large and, of course, if nature
kindly agrees to follow our theoretical pictures of these phenomena.  We
propose several new names for sources previously classified as ``radio
relics'' and suggest, with no small hint of hubris, that all who read this
article start using them.

We have attempted to classify the sources discussed herein by their current
physical interpretations rather than by their phenomenological properties,
as this promises to establish a firmer basis for our classifications.
While some of these physical interpretations may need to be modified or
re-worked entirely as the quality of the data improves, we view this as a
natural component of any scientific endeavor.  It is even possible that
some of the sources we identify here will ultimately require new, as yet
unimagined physical interpretations.  If some of the sources we mention
need to be re-classified in future, so be it.  This article lays the
groundwork for a rationalized classification scheme for cluster radio
sources, and we expect (and hope!) that this scheme will be built upon in
the future.

\begin{deluxetable*}{lllccll}
\tablecaption{Summary of properties of cluster radio sources
\label{Confnote:properties}}
\tablehead{
\colhead{} &
\multicolumn{4}{c}{Radio Source Characteristics} &
\colhead{Relationship} &
\colhead{} \\ \cline{2-5}
\colhead{Type} &
\colhead{Size} &
\colhead{Morphology} &
\colhead{$\alpha$} &
\colhead{Polarization} &
\colhead{to Hot Gas} &
\colhead{Prototype}
}
\startdata
\multicolumn{7}{l}{\bf Associated with active radio galaxies:}\\
VLBI Core                        & 10--100 pc                  & multiple point sources                                                                         & $-0.5$ & few \%        & None            & 3C~345  \\
Confined Cluster Core Source     & 10 kpc                      & \parbox[t][3em][t]{1.4in}{core + halo that may or may not include distinct lobes}              & $\lesssim-1.5$ & $\lesssim60$\%\tablenotemark{\dag$\star$} & Anti-correlated & Perseus \\
Radio Galaxy                     & few$\times10^2$ kpc         & \parbox[t][2em][t]{1.4in}{core + jets + outer lobes, possibly misaligned}                      &$<-0.6$\tablenotemark{\dag} &few$\times10$\%\tablenotemark{\dag}& \parbox[t][2em][t]{0.65in}{May be anti-correlated} & Hydra A \\
Classical Double\tablenotemark{*}& few$\times10^2$--$10^3$ kpc & core + jets                                                                                    &$<-0.6$\tablenotemark{\dag} &few$\times10$\%\tablenotemark{\dag}& \parbox[t][2em][t]{0.65in}{May be anti-correlated} & Cygnus A \\
\multicolumn{7}{l}{\bf Associated with extinct/dying radio galaxies:}\\
AGN Relic\tablenotemark{\ddag}   & few$\times10$ kpc           & \parbox[t][3em][t]{1.4in}{filamentary + some diffuse emission; more extended at low frequency} & $\lesssim-1.5$ & $\lesssim20$\% & \parbox[t][2em][t]{0.65in}{May be anti-correlated} & A133    \\
Phoenix\tablenotemark{\ddag}     & $10^2$ kpc                  & \parbox[t][3em][t]{1.4in}{filamentary + some diffuse emission; more extended at low frequency} & $\lesssim-1.5$ & 10--30\%     & \parbox[t][3em][t]{0.65in}{merger/ accretion shocks} & A85     \\
\multicolumn{7}{l}{\bf Not associated with radio galaxies:}\\
Radio Gischt\tablenotemark{\ddag}& few$\times10^2$--$10^3$ kpc & \parbox[t][3em][t]{1.4in}{possible filaments, mostly diffuse; often two symmetric sources}    & $\lesssim-1.2$ & 10--30\% lin. & Merger shocks   & A3667   \\
Mini-Halo                        & few$\times10^2$ kpc         & diffuse, centrally peaked                                                                      & $< -1.5$ & $\lesssim$ few\% & Correlated      & Perseus \\
Halo                             & $\gtrsim10^3$ kpc           & \parbox[t][3em][t]{1.4in}{diffuse, centrally peaked, may be asymmetric, may have substructure} & $\lesssim-1.1$ & none         & Correlated      & 1E0657-56 \\
\enddata
\tablenotetext{*}{Rare at low redshift.}
\tablenotetext{\dag}{Frequency dependent.}
\tablenotetext{$\star$}{Variable across source.}
\tablenotetext{\ddag}{Often called ``radio relic'' in the literature.}
\end{deluxetable*}

\section{Classifications}
\label{Confnote:class}

The numerous types of large-scale radio sources in clusters of galaxies
span approximately 5 orders of magnitude in linear size and possess a wide
range of radio properties.  They can be broken down into two basic classes:
those associated with AGN, and those associated with the ICM.  The first of
these two classes is the larger of the two, comprising all but three of the
source types we will discuss here.  This first class can be broken down
further into sources associated with currently active AGN and those
associated with extinct or dying AGN.  Of the three types not associated
with AGN, two appear to be associated with cluster mergers while one, the
``mini-halo,'' has an even less certain origin and may not even be a
distinct class of source.  Here, then, are the 9 classes of extended
cluster radio sources with their fundamental observational properties and
generally agreed upon (though occasionally uncertain) physical origins.  We
have arranged them roughly in order of increasing linear size.  The basic
properties of these sources are summarized in
Table~\ref{Confnote:properties}.

\subsection{Sources Associated With Active AGN}

{\bf VLBI Core:} The smallest of the sources discussed here, these
small-scale jets are unresolved in single-dish and small array
observations.  At the milli-arcsecond resolution of VLBI they are resolved
into multiple knots of emission, sometimes with accompanying diffuse
emission, in one- or two-sided jets plus a core associated with the central
black hole.  Typical size scales of these jets are ten to a hundred
parsecs.  Proper motions of the knots are observed over timescales of
several months, with inferred superluminal motion \citep[for an early
review, see][]{cu82}.  Circular polarization at the level of $\lesssim 1$\%
is observed in some sources \citep{hom00}, while linear polarization at the
level of a few percent is also detected in some sources
\citetext{\citealp{ens03} and references therein}.  The spectrum of the
black hole core is usually flat, while the jets have typical spectral
indices of $\alpha \sim -0.5$, where S$_\nu \propto \nu^\alpha$.  These
sources are often observed in the cores of larger scale radio sources,
e.g. M87 \citep{sr85}.

\smallskip

{\bf Confined Cluster Core (CCC) Source:} These small ($\sim$10 kpc)
sources have been the subject of much study recently with the advent of
X-ray observations with high angular resolution.  The sort of ``bubbles''
discovered by {\it ROSAT} in the core of the Perseus cluster have now been
seen with {\it Chandra} in many cooling flow clusters \citep[and
others]{fse+00,bsm+01}.  Unlike the more extended FR~I radio galaxies,
these sources often do not show two distinct radio lobes.  While individual
lobes may be visible, they are usually embedded in a halo of diffuse radio
emission that extends in all azimuthal angles around the source center.
When visible, the radio lobes are often quite distorted, as in the
Centaurus cluster \citep*{tfa02}.  These sources are typically found at the
centers of cooling flow clusters, where the high density of the ICM
confines the radio plasma, creating the amorphous morphology of these
sources.  As the radio sources interact strongly with the ICM, they inflate
large holes or ``bubbles'' in the hot gas.  Bubbles from previous epochs of
AGN outburst are occasionally seen at larger distances from the center of
the cluster and are sometimes filled with older radio plasma, as is the
case with at least one of the outer bubbles in the Perseus cluster
\citep{fse+00}.  It should be noted that these are not the only radio
sources that inflate bubbles in the ICM.

These sources should not be confused with classical FR~I radio galaxies
which have much more distinct jets and radio lobes and are generally larger
in physical extent.  CCC sources may or may not have clearly
distinguishable jets, even when more distinct lobes are visible.  For
example, the prototypical CCC source, 3C~84 at the center of the Perseus
cluster, has visible jets \citep{lz02}, while another well studied source,
A2052, does not \citep{bsm+01} They do, however, frequently have a bright,
distinct core.  As mentioned above, CCC sources are generally found in
cooling flow clusters and not in less relaxed clusters where the ICM is
less dense and may not be stable for long enough to enable X-ray bubbles to
be inflated.

\smallskip

{\bf Radio Galaxy (FR~I, WAT, and NAT):} Unlike the previous class of
sources, these classical radio galaxies have well defined jets with
expanded envelopes, or lobes, at larger radii.  Hydra A is a perfect
example of a classical FR~I.  The symmetric jets typically extend on the
order of 10 kpc before the outer envelope forms.  Outside this radius, the
jets expand into radio lobes that are typically anti-correlated with the
X-ray emission \citep{mwn+00}.  It is not clear at this time whether the
separation of the outer envelope from the jets is caused by a drop in the
density of the ICM or vice versa.  The X-ray bubbles created by these
sources are seen at a greater distance from the cluster center than are the
bubbles created by confined cluster core sources.

At larger radii, some (and maybe all, if imaged with enough sensitivity)
FR~I sources show faint, extended emission beyond the brighter radio lobes.
For example, deep radio observations of Hydra A have revealed much more
extended radio emission beyond the well studied inner lobes
\citep{lct+03,tay03}.  This emission is typically misaligned with the inner
jets and lobes.  An extreme example of this misalignment is seen in A2029,
where the outer emission extends almost perpendicular to the inner lobes
\citep*{tbg94}.  For this reason, this source may also be categorized as a
wide angle tail (WAT) source.  Unlike the inner radio lobes, these outer
lobes are not necessarily anti-correlated with the X-ray emission.  The
bending of the lobes in WATs is believed to be caused by motion of the host
galaxy relative to the ICM in which it is embedded.  Some radio galaxies
with particularly large velocities relative to their host clusters have
their jets bent so much that they merge into a single extended tail behind
the galaxy.  These sources are known as narrow angle tails, or NATs.

\smallskip

{\bf Classical Double or FR~II:} The prototypical classical double radio
galaxy is Cygnus A, appearing fairly linear and symmetric at low
resolution, with bright hot spot regions near its termini.  Unlike Cygnus
A, most classical doubles are not found in rich clusters.  These radio
galaxies typically have weak jets, if any, and often only one is visible,
perhaps because of relativistic beaming.  They have sizes of order 100 kpc,
higher luminosities than the other types of sources discussed here
\citetext{$P_{1.4\, {\rm GHz}} > 10^{25}$ W~Hz$^{-1}$; \citealp{lo96}} and
are associated with giant elliptical galaxies.  Their bright bridges of
emission, connecting the hot spot regions with the core, have progressively
steeper radio spectra toward the core, commonly interpreted as being due to
radiative aging of the relativistic electrons as these flow back from the
shocks in the hot spots.  The term FR~II (Fanaroff and Riley type II, hot
spots greater than one half distance to the terminus) is often used to
denote these sources.

\subsection{Sources Associated With Extinct or Dying AGN}

{\bf AGN Relic:} Of the several types of sources commonly referred to as
``radio relics,'' sources associated with extinct or dying AGN are the only
ones where the name is appropriate.  We propose to reserve the term
``relics'' for these objects that are indeed relic radio galaxies where the
AGN outburst that created the radio lobes has switched off, leaving the
radio plasma to passively evolve.  Because of the relatively short
radiative lifetimes of the electrons in radio lobes, these sources are
necessarily found near their source galaxies, in the central 10s of kpc of
the cluster.  These sources are often quite filamentary when observed with
sufficient angular resolution, as is the case with the relic in A133
\citep{srm+01}.  While the relic in A133 does not at first appear to be an
obvious relic radio lobe, an analysis of the {\it Chandra} observation of
the cluster showed this to be the most likely scenario \citep{fsk+02}.  A
similar source has been found in MKW~3s, where a cold, bright finger of
X-ray emission leads to a filamentary radio source \citep{sra98,mkp+02}.
Recent multi-wavelength observations by Young and collaborators have
highlighted the radio bridge connecting the filamentary source back to an
AGN at the center of the cluster \citep{you03}.

As AGN relics age, their electrons will lose so much energy that their
momentum spectra will eventually steepen to the point where they do not
emit significant radiation at high frequencies and will appear simply as
X-ray cavities with no observable radio emission.  A theoretical discussion
of these ``radio ghosts'' was outlined by \citet{ens99}.  At later times,
the emission may only be observable at frequencies below the limits of
today's instruments.  The planned low-frequency observatories
LOFAR\footnote{http://www.lofar.org} and
SKA\footnote{http://www.skatelescope.org} may be able to detect radio
emission from these ghost cavities where current instruments fail.

\smallskip

{\bf Radio Phoenix:} Previously known as radio relics, this class of object
has one foot in each of two of our larger classes.  With their origin as
radio galaxies, these sources are clearly related to AGN, but they would
not exist without being acted upon by the ICM.  These sources begin their
lives as true relics, but with a population of electrons that have aged so
much as to be no longer emitting synchrotron radiation at currently
observable radio frequencies (although LOFAR and/or the SKA may make such
sources visible).  When a merger shock or accretion shock from a cluster
merger passes through such a faded relic, it will compress the fossil radio
plasma.  Because the density in the fossil plasma is expected to be much
lower than that in the surrounding ICM, these shocks may become sub-sonic
compression waves inside the plasma, as shown by \citet{eg01} and
\citet{eb02}.  They also showed that these compression waves are capable of
re-accelerating the electrons in the fossil plasma to energies at which
they will once again be visible in the radio.  The fossil plasma is thereby
``reborn'' from its own ashes to become a radio phoenix.

It is possible to use radio observations performed at at least three
different frequencies to determine the original spectral index of a radio
phoenix.  In this technique, color-color data is fit with a theoretical
model for the spectrum of an aged electron population.  Given the current
spectral shape, the original spectral slope is uniquely determined under
the assumption of a relativistic electron population with a constant
spectral shape over the entire source.  For example, this has been done for
the steep spectrum source associated with the southwest subcluster in A85.
An original spectral index of $-0.85$ was found, confirming its likely
origin as fossil radio plasma from an extinct AGN \citep{you03}.

\subsection{Sources Associated With the ICM}

{\bf Radio Gischt:} For many years, these sources have been identified as
``radio relics.''  The name ``radio flotsam'' has also been proposed for
these sources, but is not in wide use.  We propose the name ``radio
gischt,'' German for the froth or spray on the tops of ocean waves.  This
name derives from the physical interpretation of these sources as
synchrotron radiation from electrons directly accelerated from the thermal
plasma in shocks.  This name may seem somewhat similar to ``radio
flotsam,'' but it more explicitly references the connection between the
radio sources and the pressure waves (merger/accretion shocks) responsible
for them.

As we already mentioned, these sources are believed to be synchrotron
emission from electrons directly accelerated from the thermal plasma in
merger/accretion shocks.  The radiative lifetime of $\gamma \sim 10^{4-5}$
electrons is only on the order of $10^8$ years, comparable to the sound
crossing time of these radio sources, so the radio emitting electrons
should roughly trace out the shocks themselves.  Because these sources tend
to be found far from the centers of clusters where the X-ray surface
brightness is very low, no shocks have yet been detected in X-rays at the
positions of radio gischt.  However, \citet{mv01} found an enhancement of
the relativistic electron population relative to the thermal population at
the location of a merger shock near the center of A665.  This provides the
most concrete evidence thus far of in situ acceleration of electrons to
relativistic energies.

Since these sources are directly associated with merger shocks, they should
often come in pairs, at least in the intermediate stages of mergers between
clusters with relatively equal masses.  These should be located on opposite
sides of the cluster along the axis of the merger, with the individual
radio structures elongated perpendicular to this axis.  Only a few
symmetric gischt have been seen to date, in A3667 \citep{rwh+97,jce+02},
A3376 \citep{mur99,bag02}, and A1240 \citep{ks01}, and the
last of these has yet to be confirmed by deep pointed observations.

Their spectra are quite steep ($\alpha \sim -1.2$), and they are frequently
found to be linearly polarized at the level of about 10\%--30\%
\citep[e.g.\ A2256,][]{ce01}.  They are generally found on the peripheries
of clusters.  Projection effects should reduce this tendency somewhat, as
may be the case in A2256.  The absence of other centrally located gischt
can be explained if the sources are highly planar and thus have surface
brightnesses too low to be seen if they are observed face-on.

\smallskip

{\bf Mini-Halo:} Unlike radio gischt and halos (see below), which appear to
be tied to on-going merger events in clusters, mini-halos sources are
typically found at the centers of so-called cooling flow clusters.  They
range in size from a few hundred kpc to half a Mpc.  Mini-halos are low
surface brightness, steep spectral index regions of diffuse emission which
are found around powerful radio galaxies at the cores of some clusters
\citetext{e.g.\ PKS~0745-191: \citealp{bo91}; Perseus: \citealp{bsg+92};
Virgo: \citealp{oek00}; A2626: \citealp{rlb+00}}.  These sources are
expected to have very short radiative lifetimes due to the high magnetic
fields present in cooling-flow cores \citep{tfa02} and thus the particles
must be re-accelerated or injected {\it in situ}.  Unlike radio gischt and
halos, however, these sources probably do not draw their energy from major
merger events as there is an anti-correlation between the presence of large
mergers and that of cooling-flows.

\citet*{gbs02} propose that the energy required for the particle
acceleration comes from the cooling flow.  They suggest a model where the
electrons are continually undergoing re-acceleration due to the MHD
turbulence associated with the cooling-flow region.  This model produces
Fermi-type acceleration which would result in a particle population that
steepens away from the cluster center.  This seems to be in general
agreement with observations of the Perseus mini-halo which shows spectral
index steepening away from the cluster core \citep{sij93} although some
evidence suggests this steepening may be an observational artifact
\citep*{epv03,pv03,pe03}.  \citet{gbs02} suggest that the mini-halos are
rare due to the need for a careful balance in the ICM: there must be
sufficient turbulence to balance the radiative losses of the particles but
not so much turbulence that it disrupts the cooling-flow.

A model for the formation of mini-halos, in which the relativistic
electrons are accelerated by MHD turbulence in cooling flows, is presented
by \citet{gbs+03} in the proceedings of this conference.

\citet{pe03}, on the other hand, propose a hadronic origin of these radio
mini-halos: hadronic interactions of cosmic ray protons with ambient
thermal protons produce synchrotron emitting relativistic electrons.
Azimuthally averaged radio surface brightness profiles of the Perseus
mini-halo matches the expected emission by this hadronic scenario well on
all scales.  Moreover, the small amount of required energy density in
cosmic ray protons ($\sim$2\% relative to the thermal energy density)
supports this hypothesis not only because cosmological simulations carried
out by \citet{mrk+01} easily predict a population old cosmic ray protons at
the clusters center of this order of magnitude.

\smallskip

{\bf Radio Halo:} Like radio gischt, these sources are also believed to be
the products of mergers between roughly equal mass clusters.  Thus far,
they have only been found in clusters currently undergoing or on the tail
end of a merger.  Their radio power appears to be strongly correlated with
the X-ray luminosity of the host cluster \citep{lhb+00,fer00}, although the
observational selection effects that may contribute to this relation have
not yet been fully explored.  \citet{lhb+00} also suggest a relation
between radio power and X-ray temperature, although this relation has a
much larger scatter and can be understood as simply a combination of the
$P_\nu$--$L_X$ relation and the well known $L_X$--$T_X$ relation.

The current leading hypothesis for the origin of the relativistic electrons
that create these sources is that they are re-accelerated, probably by
turbulence, from a population of relatively low energy ($\gamma \sim 10^3$)
electrons.  This mildly relativistic pool may in turn have its origin in
the merger shocks that cause radio gischt.  As mentioned above, electrons
with $\gamma \sim 10^{4-5}$ have very short lifetimes whereas $\gamma \sim
10^3$ electrons have lifetimes of order a few times $10^9$ years, so any
electrons accelerated in gischt would be capable of providing this
supra-thermal pool later in the merger while not actively emitting
synchrotron radiation in the absence of a mechanism of re-acceleration.
\citet*{fts03} demonstrated that turbulence may be capable of this
re-acceleration in the case of roughly equal mergers between high mass
systems.  Turbulence has the attractive feature that it is directly
connected to the local properties of the ICM and therefore should create
radio emission that is correlated with local variations in the ICM.  This
connection has been demonstrated in a number of clusters, where strong
correlations have been measured between the surface brightness distribution
of radio halos and their host clusters' X-ray emission
\citep{gef+01,gfg+01}.

Another scenario, less favored at the moment but still plausible, is
so-called secondary electron model proposed originally by \citet{den80} and
examined in more detail by \citet{bc99} and \citet{de00}.  This model
produces the radio-emitting electrons as a byproduct of hadronic
interactions between cosmic ray protons (CRp's) and and the ambient thermal
gas.  Because they have lifetimes greater than a Hubble time, CRp's from a
central AGN can diffuse throughout a cluster, producing the observed large
scale emission.  The steep falloff of the emission predicted by these
models, however, does not match the observations of halos
\citep{de00,gef+01,bru02} but may still fit those of mini-halos (see
above and \citealp{epv03,pe03}).

Spectra of these sources are similar to those of gischt, though slightly
flatter ($\alpha \sim -1.1$).  Unlike gischt, however, radio halos are
completely unpolarized.


\acknowledgements
We extend a special thanks to all those who participated in the discussions
at the conference about the naming and classification of these sources.  We
also thank Christoph Pfrommer and Corina Vogt for their proofreading of and
helpful comments on the manuscript.  Finally, we also thank Thomas Reiprich
for producing one of the best organized conference we have ever had the
pleasure to attend.

\bibliography{xray,radio}

\begin{thebibliography}{48}
\expandafter\ifx\csname natexlab\endcsname\relax\def\natexlab#1{#1}\fi

\bibitem[{{Bagchi}(2002)}]{bag02}
{Bagchi}, J. 2002, \mnras, submitted (astro-ph/0210553)

\bibitem[{{Baum} \& {O'Dea}(1991)}]{bo91}
{Baum}, S.~A. \& {O'Dea}, C.~P. 1991, \mnras, 250, 737

\bibitem[{{Blanton} {et~al.}(2001){Blanton}, {Sarazin}, {McNamara}, \&
  {Wise}}]{bsm+01}
{Blanton}, E.~L., {Sarazin}, C.~L., {McNamara}, B.~R., \& {Wise}, M.~W. 2001,
  \apjl, 558, L15

\bibitem[{{Blasi} \& {Colafrancesco}(1999)}]{bc99}
{Blasi}, P. \& {Colafrancesco}, S. 1999, Astroparticle Physics, 12, 169

\bibitem[{{Brunetti}(2002)}]{bru02}
{Brunetti}, G. 2002, in Matter and Energy in Clusters of Galaxies, ed.
  S.~{Bowyer} \& C.-Y. {Hwang} (ASP Conference Series), in preparation
  (astro-ph/0208074)

\bibitem[{{Burns} {et~al.}(1992){Burns}, {Sulkanen}, {Gisler}, \&
  {Perley}}]{bsg+92}
{Burns}, J.~O., {Sulkanen}, M.~E., {Gisler}, G.~R., \& {Perley}, R.~A. 1992,
  \apjl, 388, L49

\bibitem[{{Clarke} \& {Ensslin}(2001)}]{ce01}
{Clarke}, T.~E. \& {Ensslin}, T.~A. 2001, in {Proceedings of XXI Moriond
  Astrophysics Meeting: Galaxy Clusters and the High Redshift Universe Observed
  in X-rays}, ed. D.~{Neumann}, F.~{Durret}, , \& J.~{Tran Thanh Van},
  (astro-ph/0106137)

\bibitem[{{Cohen} \& {Unwin}(1982)}]{cu82}
{Cohen}, M.~H. \& {Unwin}, S.~C. 1982, in IAU Symp. 97: Extragalactic Radio
  Sources, 345

\bibitem[{{Dennison}(1980)}]{den80}
{Dennison}, B. 1980, \apjl, 239, L93

\bibitem[{{Dolag} \& {En{\ss}lin}(2000)}]{de00}
{Dolag}, K. \& {En{\ss}lin}, T.~A. 2000, \aap, 362, 151

\bibitem[{{En{\ss}lin}(1999)}]{ens99}
{En{\ss}lin}, T.~A. 1999, in Diffuse Thermal and Relativistic Plasma in Galaxy
  Clusters, 275

\bibitem[{{En{\ss}lin}(2003)}]{ens03}
{En{\ss}lin}, T.~A. 2003, \aap, 401, 499

\bibitem[{{En{\ss}lin} \& {Br{\" u}ggen}(2002)}]{eb02}
{En{\ss}lin}, T.~A. \& {Br{\" u}ggen}, M. 2002, \mnras, 331, 1011

\bibitem[{{En{\ss}lin} \& {Gopal-Krishna}(2001)}]{eg01}
{En{\ss}lin}, T.~A. \& {Gopal-Krishna}. 2001, \aap, 366, 26

\bibitem[{{En{\ss}lin} {et~al.}(2003){En{\ss}lin}, {Pfrommer}, \&
  {Vogt}}]{epv03}
{En{\ss}lin}, T.~A., {Pfrommer}, C., \& {Vogt}, C. 2003, these proceedings

\bibitem[{{Fabian} {et~al.}(2000){Fabian}, {Sanders}, {Ettori}, {Taylor},
  {Allen}, {Crawford}, {Iwasawa}, {Johnstone}, \& {Ogle}}]{fse+00}
{Fabian}, A.~C., {Sanders}, J.~S., {Ettori}, S., {Taylor}, G.~B., {Allen},
  S.~W., {Crawford}, C.~S., {Iwasawa}, K., {Johnstone}, R.~M., \& {Ogle}, P.~M.
  2000, \mnras, 318, L65

\bibitem[{{Feretti}(2000)}]{fer00}
{Feretti}, L. 2000, in The Universe at Low Radio Frequencies, ed. G.~{Swarup}
  \& A.~{Pramesh Rao} (San Francisco: Astr. Soc. Pacific), in press
  (astro-ph/0006379)

\bibitem[{{Fujita} {et~al.}(2002){Fujita}, {Sarazin}, {Kempner}, {Rudnick},
  {Slee}, {Roy}, {Andernach}, \& {Ehle}}]{fsk+02}
{Fujita}, Y., {Sarazin}, C.~L., {Kempner}, J.~C., {Rudnick}, L., {Slee}, O.~B.,
  {Roy}, A.~L., {Andernach}, H., \& {Ehle}, M. 2002, \apj, {757}, 764

\bibitem[{{Fujita} {et~al.}(2003){Fujita}, {Takizawa}, \& {Sarazin}}]{fts03}
{Fujita}, Y., {Takizawa}, M., \& {Sarazin}, C.~L. 2003, \apj, {584}, 190

\bibitem[{{Gitti} {et~al.}(2002){Gitti}, {Brunetti}, \& {Setti}}]{gbs02}
{Gitti}, M., {Brunetti}, G., \& {Setti}, G. 2002, \aap, 386, 456

\bibitem[{{Gitti} {et~al.}(2003){Gitti}, {Brunetti}, {Setti}, \&
  {Feretti}}]{gbs+03}
{Gitti}, M., {Brunetti}, G., {Setti}, G., \& {Feretti}, L. 2003, these proceedings

\bibitem[{{Govoni} {et~al.}(2001{\natexlab{a}}){Govoni}, {En{\ss}lin},
  {Feretti}, \& {Giovannini}}]{gef+01}
{Govoni}, F., {En{\ss}lin}, T.~A., {Feretti}, L., \& {Giovannini}, G.
  2001{\natexlab{a}}, \aap, 369, 441

\bibitem[{{Govoni} {et~al.}(2001{\natexlab{b}}){Govoni}, {Feretti},
  {Giovannini}, {B{\" o}hringer}, {Reiprich}, \& {Murgia}}]{gfg+01}
{Govoni}, F., {Feretti}, L., {Giovannini}, G., {B{\" o}hringer}, H.,
  {Reiprich}, T.~H., \& {Murgia}, M. 2001{\natexlab{b}}, \aap, 376, 803

\bibitem[{{Homan}(2000)}]{hom00}
{Homan}, D.~C. 2000, Ph.D.~Thesis

\bibitem[{{Johnston-Hollitt} {et~al.}(2002){Johnston-Hollitt}, {Clay}, {Ekers},
  {Wieringa}, \& {Hunstead}}]{jce+02}
{Johnston-Hollitt}, M., {Clay}, R.~W., {Ekers}, R.~D., {Wieringa}, M.~H., \&
  {Hunstead}, R.~W. 2002, in IAU Symposium, 157

\bibitem[{{Kempner} \& {Sarazin}(2001)}]{ks01}
{Kempner}, J.~C. \& {Sarazin}, C.~L. 2001, \apj, 548, 639

\bibitem[{Lane {et~al.}(2003)Lane, Clarke, Taylor, Perley, \& Kassim}]{lct+03}
Lane, W., Clarke, T.~E., Taylor, G.~B., Perley, R.~A., \& Kassim, N.~E. 2003,
  \apj, submitted

\bibitem[{{Ledlow} \& {Owen}(1996)}]{lo96}
{Ledlow}, M.~J. \& {Owen}, F.~N. 1996, \aj, 112, 9

\bibitem[{{Liang} {et~al.}(2000){Liang}, {Hunstead}, {Birkinshaw}, \&
  {Andreani}}]{lhb+00}
{Liang}, H., {Hunstead}, R.~W., {Birkinshaw}, M., \& {Andreani}, P. 2000, \apj,
  544, 686

\bibitem[{{Liu} \& {Zhang}(2002)}]{lz02}
{Liu}, Y. \& {Zhang}, H.~Q. 2002, \aap, 386, 646

\bibitem[{{Markevitch} \& {Vikhlinin}(2001)}]{mv01}
{Markevitch}, M. \& {Vikhlinin}, A. 2001, \apj, 563, 95

\bibitem[{{Mazzotta} {et~al.}(2002){Mazzotta}, {Kaastra}, {Paerels},
  {Ferrigno}, {Colafrancesco}, {Mewe}, \& {Forman}}]{mkp+02}
{Mazzotta}, P., {Kaastra}, J.~S., {Paerels}, F.~B., {Ferrigno}, C.,
  {Colafrancesco}, S., {Mewe}, R., \& {Forman}, W.~R. 2002, \apjl, 567, L37

\bibitem[{{McNamara} {et~al.}(2000){McNamara}, {Wise}, {Nulsen}, {David},
  {Sarazin}, {Bautz}, {Markevitch}, {Vikhlinin}, {Forman}, {Jones}, \&
  {Harris}}]{mwn+00}
{McNamara}, B.~R., {Wise}, M., {Nulsen}, P.~E.~J., {David}, L.~P., {Sarazin},
  C.~L., {Bautz}, M., {Markevitch}, M., {Vikhlinin}, A., {Forman}, W.~R.,
  {Jones}, C., \& {Harris}, D.~E. 2000, \apjl, 534, L135

\bibitem[{{Miniati} {et~al.}(2001){Miniati}, {Ryu}, {Kang}, \&
  {Jones}}]{mrk+01}
{Miniati}, F., {Ryu}, D., {Kang}, H., \& {Jones}, T.~W. 2001, \apj, 559, 59

\bibitem[{{Murphy}(1999)}]{mur99}
{Murphy}, T. 1999, {Honours Thesis, University of Sydney}

\bibitem[{{Owen} {et~al.}(2000){Owen}, {Eilek}, \& {Kassim}}]{oek00}
{Owen}, F.~N., {Eilek}, J.~A., \& {Kassim}, N.~E. 2000, \apj, 543, 611

\bibitem[{{Pfrommer} \& {En{\ss}lin}(2003)}]{pe03}
{Pfrommer}, C. \& {En{\ss}lin}, T.~A. 2003, \aap, 407, L73

\bibitem[{{Pfrommer} \& {Vogt}(2003)}]{pv03}
{Pfrommer}, C. \& {Vogt}, C. 2003, these proceedings

\bibitem[{{Rizza} {et~al.}(2000){Rizza}, {Loken}, {Bliton}, {Roettiger},
  {Burns}, \& {Owen}}]{rlb+00}
{Rizza}, E., {Loken}, C., {Bliton}, M., {Roettiger}, K., {Burns}, J.~O., \&
  {Owen}, F.~N. 2000, \aj, 119, 21

\bibitem[{{R\"ottgering} {et~al.}(1997){R\"ottgering}, {Wieringa}, {Hunstead},
  \& {Ekers}}]{rwh+97}
{R\"ottgering}, H.~J.~A., {Wieringa}, M.~H., {Hunstead}, R.~W., \& {Ekers},
  R.~D. 1997, \mnras, 290, 577

\bibitem[{{Schmitt} \& {Reid}(1985)}]{sr85}
{Schmitt}, J.~H.~M.~M. \& {Reid}, M.~J. 1985, \apj, 289, 120

\bibitem[{{Sijbring}(1993)}]{sij93}
{Sijbring}, D. 1993, PhD thesis, Gr\"{o}ningen

\bibitem[{{Slee} {et~al.}(1998){Slee}, {Roy}, \& {Andernach}}]{sra98}
{Slee}, O.~B., {Roy}, A.~L., \& {Andernach}, H. 1998, Australian Journal of
  Physics, 51, 971

\bibitem[{{Slee} {et~al.}(2001){Slee}, {Roy}, {Murgia}, {Andernach}, \&
  {Ehle}}]{srm+01}
{Slee}, O.~B., {Roy}, A.~L., {Murgia}, M., {Andernach}, H., \& {Ehle}, M. 2001,
  \aj, 122, 1172

\bibitem[{{Taylor}(2003)}]{tay03}
{Taylor}, G.~B. 2003, this volume

\bibitem[{{Taylor} {et~al.}(1994){Taylor}, {Barton}, \& {Ge}}]{tbg94}
{Taylor}, G.~B., {Barton}, E.~J., \& {Ge}, J. 1994, \aj, 107, 1942

\bibitem[{{Taylor} {et~al.}(2002){Taylor}, {Fabian}, \& {Allen}}]{tfa02}
{Taylor}, G.~B., {Fabian}, A.~C., \& {Allen}, S.~W. 2002, \mnras, 334, 769

\bibitem[{{Young}(2003)}]{you03}
{Young}, A. 2003, PhD thesis, Univ.~Minnesota

\end{thebibliography}
\bibliographystyle{apj}

\end{document}